\pdfoutput=1

\documentclass[sigconf,nonacm]{acmart}

\usepackage{graphicx}
\usepackage{hyperref}
\usepackage{color}
\usepackage{booktabs}
\usepackage{amsmath}
\usepackage{algorithm}
\usepackage[noend]{algpseudocode}

\AtBeginDocument{%
  }

\begin{document}

\title{Building an Explainable Graph-based Biomedical Paper Recommendation System (Technical Report)}

\author{Hermann Kroll}
\email{krollh@acm.org}
\orcid{0000-0001-9887-9276}
\affiliation{%
  \institution{Institute for Information Systems,\\ TU Braunschweig}
\country{Germany}
}

\author{Christin K. Kreutz}
\email{ckreutz@acm.org}
\orcid{0000-0002-5075-7699}
\affiliation{%
  \institution{TH Mittelhessen \& Herder Institute}
  \country{Germany}
}

\author{Bill Matthias Thang}
\email{m.thang@tu-bs.de}
\orcid{0009-0006-8321-8479}
\affiliation{%
  \institution{Institute for Information Systems,\\ TU Braunschweig}
    \country{Germany}
}

\author{Philipp Schaer}
\email{philipp.schaer@th-koeln.de}
\orcid{0000-0002-8817-4632}
\affiliation{%
  \institution{TH Köln (University of Applied Sciences)}
    \country{Germany}
}

\author{Wolf-Tilo Balke}
\email{balke@ifis.cs.tu-bs.de}
\orcid{0000-0002-5443-1215}
\affiliation{%
  \institution{Institute for Information Systems,\\ TU Braunschweig}
    \country{Germany}
}

\renewcommand{\shortauthors}{Kroll et al.}

\begin{abstract}
Digital libraries provide different access paths, allowing users to explore their collections. For instance, paper recommendation suggests literature similar to some selected paper. Their implementation is often cost-intensive, especially if neural methods are applied. Additionally, it is hard for users to understand or guess why a recommendation should be relevant for them. That is why we tackled the problem from a different perspective. We propose XGPRec, a graph-based and thus explainable method which we integrate into our existing graph-based biomedical discovery system. Moreover, we show that XGPRec (1) can, in terms of computational costs, manage a real digital library collection with 37M documents from the biomedical domain, (2) performs well on established test collections and concept-centric information needs, and (3) generates explanations that proved to be beneficial in a preliminary user study. We share our code so that user libraries can build upon XGPRec.
  
\end{abstract}



\maketitle

\section{Introduction}
\label{sec:intro}
Digital libraries provide effective access paths for users to explore their underlying collections. 
However, the number of scientific publications in such collections is increasing rapidly.
This vast amount of options to look at combined with possibly under-specified information needs of users can lead to challenges when trying to find related work for a topic of interest as keyword-based options are insufficient~\cite{DBLP:journals/jodl/KreutzS22}.
As a remedy, paper recommendation systems suggest related literature based on some user's selected article(s)~\cite{DBLP:conf/jcdl/CollinsB19,DBLP:journals/corr/abs-2103-08819}.
The advantage of paper recommendation is that users get an impression of what other works are highly related to some given article, allowing them to explore a library's collection in this way.
We consider the following definition for the paper recommendation task: \textit{Given an initial article, what are other relevant articles with regard to the input article's topic and content?}

Current paper recommendation systems do not focus on limiting computational complexity or performing the task with fewer resources, even though the emissions produced by neural retrieval methods are several orders of magnitude higher than those of BM25~\cite{DBLP:conf/sigir/ScellsZZ22}. 
Moreover, even worse, current approaches often rely on deep learning~\cite{DBLP:journals/jodl/KreutzS22}, i.e., a cost-intensive training or embedding step is necessary. 
Collecting training data on a large scale while also considering a variety of different information needs and topics is an exceptionally challenging task.
If training data is \textit{just} collected for a special subgroup of users, learned system will likely have a bias in the end when answering unseen information needs.
In practice, the computational costs of neural retrieval methods and the collection of training data usually hinder their implementation in a digital library.
Another understudied issue but desirable goal in paper recommendation is explainability~\cite{DBLP:journals/jodl/KreutzS22}. 
Explanations are helpful to (1) justify individual recommendations, (2) understand how a system works, and (3) distinguish good from bad recommendations via users' feedback~\cite{DBLP:conf/sigir/BalogRA19}.
In contrast to deep learning approaches that may struggle to explain their retrieved documents to users, we choose to head for an alternative: concept-centric graph-based document representations.

We tackled the two issues of high costs (in terms of retrieval and training data collection) and missing explainability by building upon graph-based document representations in cooperation with PubPharm\footnote{\url{www.pubpharm.de}}, the specialized service for Pharmacy (some of this paper's authors are part of the PubPharm team). 
In addition to keyword-based retrieval, PubPharm offers a graph-based discovery system called the Narrative Service\footnote{\url{https://narrative.pubpharm.de}} as a more sophisticated access path for users.
The Narrative Service assumes that complex, concept-centric information needs can be represented as narrative query graphs, i.e., as short stories of interest involving relevant biomedical concepts and their interactions.
The service comes with two central advantages~\cite{DBLP:journals/jodl/KrollPKKRB24}: 
First, narrative query graphs allow users to search for literature precisely by stating concept interactions explicitly.
Second, users can explore a digital library's collection using variables, e.g., structuring the literature by possible treatment options in adults with diabetes.
While the costs for harvesting graphs from texts can be quite high, we proposed effective and cost-moderate extraction workflows for digital libraries that we applied to build our and PubPharm's system~\cite{DBLP:conf/jcdl/KrollPPB22a,kroll2023ijdlIEW}.

We expand our and PubPharm's existing Narrative Service and provide more functionality to our users. We design and implement an explainable paper recommendation system (a demo video is available here\footnote{\url{https://youtu.be/oLZFCtVuQWU}}), called \textbf{XGPRec}. We will integrate XGPRec into to our graph-based biomedical discovery system~\cite{DBLP:journals/jodl/KrollPKKRB24}.
By consciously refraining from using computationally intense neural retrieval methods in our recommendation and instead relying on graph-based similarities and text-scoring through BM25, we bypass vast computational costs~\cite{DBLP:conf/sigir/ScellsZZ22}.
Graph representations can visualize complex interconnections and enable users from the biomedical domain to immediately grasp if a paper fits what they mean to find~\cite{DBLP:conf/jcdl/KrollKSB23}. 
In this work, we focus on justifying recommendations~\cite{DBLP:conf/sigir/BalogRA19}, which we achieve by displaying overlapping graph patterns between an initial and a candidate paper to explain recommendations for users.

At the showcase of a real digital library, we demonstrate how graph-based recommendation can be implemented for the extensive MEDLINE document collection, which contains about 37 million documents.
Our research objective for this work is thus:
\textit{How can we design a fast and reliable paper recommendation system?}
More precisely: \textit{(1) How can we find relevant candidate documents with as low costs as possible (first stage)?}
\textit{(2) How can we compute relevant and explainable recommendations (second stage)?}

Our code is available at GitHub\footnote{\url{https://github.com/HermannKroll/NarrativeRecommender/}} and Software Heritage\footnote{Software Heritage ID:\href{https://archive.softwareheritage.org/swh:1:dir:eaeaac5c6a9ccb00542431398e43dec34d910faf}{swh:1:dir:eaeaac5c6a9ccb00542431398e43dec34d910faf}}.

\section{Related Work}
\label{sec:related_work}

\subsection{Paper Recommendation}
Several surveys describe approaches and problems of the general paper recommendation domain~\cite{DBLP:journals/access/BaiWLYKX19,DBLP:journals/jodl/BeelGLB16,DBLP:journals/jodl/KreutzS22,DBLP:journals/eaai/SharmaGM23,DBLP:journals/kais/ZhangPYZSRCW23}. 
Our approach operates on the \textit{biomedical domain}, incorporates a \textit{knowledge graph} for every paper and is \textit{explainable}.

\textbf{Biomedical Domain.}
EILEEN~\cite{DBLP:journals/corr/abs-2110-09663} operates on a dataset from MEDLINE and lets users indicate binary relevancy of search results. Tf-idf, SVD, Elasticsearch and key phrase extraction with RAKE are combined.
ChemVis~\cite{DBLP:conf/jcdl/BreitingerHFM22} uses PubChem and depicts chemical compositions of materials while using HyPlag~\cite{DBLP:conf/sigir/MeuschkeSSG18} for general document similarity and RecVis~\cite{DBLP:conf/jcdl/BreitingerKMMG20} for recommendation.
Guo et al.~\cite{DBLP:conf/medinfo/GuoSZH21} use a linear combination of bibliographic coupling and textual similarity either computed with PMRA~\cite{DBLP:journals/bmcbi/LinW07}, BM25 or cosine similarity of tf-idf vectors. They evaluate their approach on the TREC Genomics dataset. 
Emati~\cite{DBLP:journals/biodb/KartMLKS22} works on articles which appeared in PubMed or ArXiv since 2000. Users can (dis)like papers or upload a list of articles they have cited as positive examples to train classifiers on relevant literature. The system uses BERT or tf-idf vectors of articles and their metadata and multinomial Naive Bayes to classify relevant and irrelevant papers.
KnowCOVID-19~\cite{DBLP:journals/access/OrucheGBCAZBMR21} uses different topic models on CORD-19 to identify topics described, genes explored and drugs used in papers.
They present an analysis of papers according to the Levels of Evidence Pyramid based on paper tagging by users.

Two approaches are available online:
PubMed displays similar articles to papers on their abstract page\footnote{\url{https://pubmed.ncbi.nlm.nih.gov/help\#computation-of-similar-articles}}. Word-weighting on stemmed titles, abstracts and MeSH terms as well as document length are considered when computing the neighboring recommendation candidates for an initial paper. The most similar documents for each paper are pre-computed and displayed to users to ensure fast response times.
LitSuggest~\cite{DBLP:journals/nar/AllotLCLL21} classifies unseen papers based on users' ratings of relevancy of papers per topic on PubMed. 

Therefore, from the biomedical domain, we could compare our approach against LitSuggest or PubMed. LitSuggest requires positive and negative training data provided by an user in order to recommend papers, therefore we disregard the system in our analysis to focus on easiness of use as given by PubMed.

\textbf{Knowledge Graphs.}
KERS~\cite{Afsar_Crump_Far_2021} is a feedback-based system recommending articles from categories, a user has interacted with before. The categories stem from an expert-built KB.
CV-Lattes~\cite{DBLP:conf/flairs/MagalhaesSCF15} uses users' published papers to build terms and concepts profiles (stored in a KB), weights of similarity vectors between them stem from a KB.
Li et al.~\cite{DBLP:journals/tois/LiCPR19} compose a KG-based embedding of paper metadata and learn a mapping between users' browsed and clicked recommended articles.
Manrique and Marino~\cite{manrique} compare different edge- and node- weighting schemes on directed graphs of single papers using an external KB.
Neethukrishnan and Swaraj~\cite{8117833} build personal ontologies for users from concepts of associated articles using the ACM taxonomy, WordNet and SVMs.
The overall goal of IBM PARSe~\cite{10.1093/jamiaopen/ooaa028} is identifying papers a user should read to update an existing KG. Named entities of papers are extracted and papers closest related to the existing KG are recommended.
Tang et al.~\cite{DBLP:journals/concurrency/TangLQ21} use an external KB to identify concepts in papers which they connect in a KG including users and papers.
Wang et al.~\cite{DBLP:conf/service/WangXTWX20} feed embedded paths from a user-paper-interaction KG into an LSTM. 

While previous works use information from KGs to enrich documents, we transform each document into a small KG. 
Each of these document graphs then represents the document's essential information (concept interactions).

\textbf{Explainable Systems.}
Missing explainability is one of the widespread problems in paper recommendation~\cite{DBLP:journals/jodl/KreutzS22}. 
TIGRIS~\cite{DBLP:conf/hci/BrunsVGZS15} provides a responsive graph linking keyword nodes to nodes representing papers fitting a user's interests. The graph substitutes a ranked result list.
ArXivDigest~\cite{DBLP:conf/cikm/GingstadJB20} is a living lab for providing personalized recommendations from arXiv papers with textual explanations. 
JTIE~\cite{DBLP:conf/ccscw/XieWPT020} describes explainability as an important aspect of their approach but does not describe the format of explanations.
Bakalov et al.~\cite{DBLP:conf/iui/BakalovMKSWBT13} highlight terms matching user' interests in abstracts of recommended papers.
Kangasrääsiö et al.~\cite{DBLP:conf/iui/KangasraasioGK15} do not explain recommendations but let users weight the input keywords in a visual interface, thus indirectly explaining the suggested papers.
LIMEADE~\cite{DBLP:journals/corr/abs-2003-04315} is a general purpose posthoc explanation approach producing weighted interpretable features, e.g., influential terms of recommended candidate papers.

Existing work often relies on textual features such as highlighting terms in abstracts. As users of the underlying DL have already been encountered to prefer graph over textual explanations~\cite{DBLP:conf/jcdl/KrollKSB23} we decided on focusing on an inherently explainable recommendation method producing graphs as explanation.

\begin{figure*}[t]
    \centering
    \includegraphics[width=0.8\textwidth, trim={0cm 7cm 4cm 0cm}]{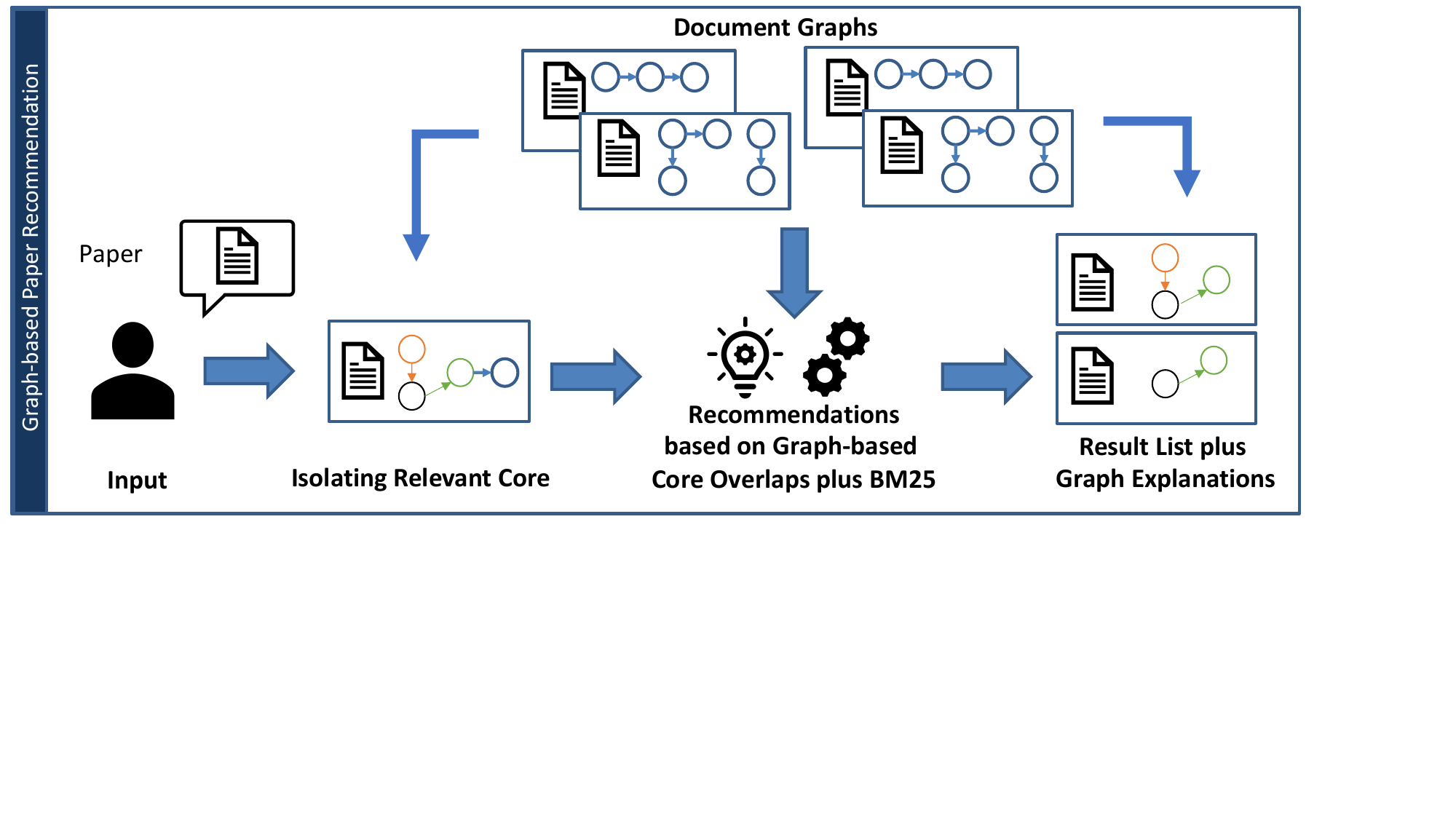}
    \caption{Systematic overview: Document graph representations are used to compute and explain relevant paper recommendations for users.}
    \Description{Systematic overview: Document graph representations are used to compute and explain relevant paper recommendations for users.}
    \label{fig:overview}
\end{figure*}

\subsection{Narrative Service - Graph-based Discovery}
\label{sec:system}
PubPharm's graph-based retrieval service, called the Narrative Service~\cite{DBLP:journals/jodl/KrollPKKRB24}, allows users to express their information need as a narrative query graph, i.e., graph patterns with triple-like statements (concept, interaction, concept). 
A query is then answered by documents that contain the search graph pattern.
For the query processing step, they transformed texts into graphs by detecting relevant concepts and extracting their interactions. 
Concepts were identified by deriving annotations from the PubTator service~\cite{wei2019pubtatorcentral,wei2013pubtator} and performing a dictionary-based concept linking through vocabularies derived from ChEBML~\cite{mendez2018chembl}, Wikidata~\cite{vrandecic2014wikidata} and the Medical Subject Headings. 
Statements (concept interactions) were extracted by using PathIE and a sentence-based extraction method that extracts general \textit{association} statements if two concepts were mentioned within the same sentence. 
Briefly, PathIE extracted a statement between two concepts if two detected concepts were connected on the dependency parse of a sentence (basically the grammatical structure) derived with the Stanford CoreNLP toolkit~\cite{DBLP:conf/acl/ManningSBFBM14}. 
Details about the extraction methods have already been published in~\cite{DBLP:conf/jcdl/KrollPB21} and~\cite{DBLP:journals/jodl/KrollPKKRB24}. 
The methods' and system's code is freely available\footnote{\url{https://github.com/HermannKroll/NarrativeIntelligence}},\footnote{Software Heritage ID:\href{https://archive.softwareheritage.org/swh:1:dir:9e2435bb03d544039cc96fa1b17537050faec6e3}{swh:1:dir:9e2435bb03d544039cc96fa1b17537050faec6e3}}. 
The discovery system currently features around 38 million publications from the MEDLINE collection and 70k COVID-19 pre-prints from ZB MED's preVIEW service~\cite{langnickel2021covid}.

\section{Graph-based Recommendation and Explanation}
The discovery system contains a set of documents $\mathcal{D}$~\cite{DBLP:journals/jodl/KrollPKKRB24}.
A document $d \in \mathcal{D}$ is represented by its text $d_{\textit{text}}$ (title and abstract), a list of concept annotations $d_{\textit{concepts}}$ and a list of extracted statements $d_{\textit{statements}}$. 
A concept annotation maps a specific text span of $d$ to a pre-known concept of the concept vocabulary $\mathcal{C}$ (the set of all known concepts by the system). 
Concepts are identified by precise IDs and a type, e.g., \textit{drug} or \textit{disease}. 
An extracted statement is composed of a subject-predicate-object triple, e.g., (\textit{Metformin}, \textit{treats}, \textit{Diabetes}), the sentence it was extracted from, and a confidence value (how good the extraction was based on some NLP method). 
Sect.~\ref{sec:system} contains details about applied methods and details.
The \textbf{document graph} of $d$ is the set of subject-predicate-object triples extracted from the document. 
A document graph is given by $\textit{graph}(d) = (V, E)$ where $V$ is the set of nodes (detected concepts) and $E$ is a set of concept interactions (triples).  
This paper aims to reuse our existing document graph representation~\cite{DBLP:journals/jodl/KrollPKKRB24} to perform a reliable and explainable paper recommendation; see Figure~\ref{fig:overview}.

\textbf{Task Definition (Explainable Paper Recommendation):} \textit{Given an input document $d$, compute a ranked list of documents $\subseteq \mathcal{D}$. For each document $d_i$ in that list provide an explanation $e_i$ of why $d_i$ is related to the input $d$.}

\subsection{Scoring Document Graph Components}
First, we score the nodes and edges of a document graph to know which parts are the most important by using three distinct features.

\textbf{Tf-idf.}
Some graph nodes or edges might carry more information (are more relevant) than very general ones. 
In information retrieval, term-frequency tf and inverse-document-frequency idf are two paradigms used to determine a term's relevance concerning a document $d$. 
We follow that paradigm and design a tf-idf score for nodes and edges. 
We define tf for a concept $c$ within a document $d$ as $\textit{tf}(c, d) = \frac{\#(c, d)}{\#C}$ with $\#(c, d)$ being the number of occurrences of concept $c$ within $d$ and $\#\mathcal{C}$ being the number of all annotated concepts within $d$ (for normalization).
Next, we define idf for a concept $c$ as $\textit{idf}(c) = \log \frac{|\mathcal{D}|}{|\{d \in \mathcal{D}\land c \in d\}|}$. 
$|\mathcal{D}|$ is the number of documents in our collection, and the denominator counts documents that include the concept $c$. 
With that, we can score nodes with tf-idf as follows:  

\begin{equation}
\textit{n-tf-idf}(n, d) = tf(n, d) \cdot \textit{idf}(n).
\end{equation}

For edges, we faced issues when maintaining an idf index:
First, the index can get quite large (quadratic growth with regard to the size of the concept vocabulary).
Second, our statement extraction methods are restricted to sentence levels and might be error-prone~\cite{DBLP:journals/jodl/KrollPKKRB24}. 
Many connections might be lost during that step, affecting the idf score.
That is why we decided to approximate the tf-idf-score for an edge by combining the tf-idf scores of its subject and object plus multiplying it with a predicate specificity (basically a score determining how specific a predicate is:  \textit{treats} is more specific than a general  \textit{association}). 
We define the tf-idf score for an edge $e = (s, p, o)$ concerning a document $d$ as: 

\begin{equation}
    \textit{e-tf-idf}(e, d) = ( \textit{n-tf-idf}(s, d) + \textit{n-tf-idf}(o, d)) \cdot \textit{specificity}(p)
\end{equation}

\textbf{Coverage.} 
The discovery system is designed for biomedical abstract retrieval.
Each abstract typically starts with some background information in the corresponding field. 
Concept mentions within that background part might be less important than concepts mentioned across the whole abstract.
We therefore define \textit{coverage} of a node (concept) $n$ within a document $d$ as:

\begin{equation}
    \textit{n-coverage}(n, d) = \frac{\textit{last\_position}(n, d) - \textit{first\_position}(n, d)}{\textit{text\_length}(d)}
\end{equation}

The method calculates the difference between the concept's last mention and the first mention within the document and normalizes it by its text length.
Coverage approximates whether the concept is used from the beginning to the end or briefly mentioned somewhere as a side note.
The higher the coverage is, the more relevant a concept $c$ should be.
We then define the coverage of a document graph's edge $e = (s, p, o)$ as:

\begin{equation}
\textit{e-coverage}(e, d) = \min(\lbrace \textit{n-coverage(s)}, \textit{n-coverage(o)}\rbrace).
\end{equation}

\textbf{Confidence.} 
As mentioned, our extraction methods come with a confidence score, i.e., a score of how sure the tool is about the corresponding extraction. 
Please note that a document graph's edge could be extracted from different sentences within $d$.
The \textbf{confidence} for an edge $e$ is defined as the maximum confidence value of the statement extractions within $d_\textit{statements}$ that support $e$.
We do not have confidence values for concept annotations because the detection methods are dictionary-based linking methods that perform a binary decision. 

\textbf{Scoring.}
Finally, we can define the scores of nodes and edges. 
Coverage and tf-idf are combined to compute the score for each node:

\begin{equation}
    \textit{n-score}(n, d) = \textit{n-coverage}(n, d) \cdot \textit{n-tf-idf}(n, d)
\end{equation}
For edges, we combine confidence, coverage, and tf-idf:
\begin{equation}
    \textit{e-score}(e, d) = \textit{confidence}(e, d) \cdot \textit{e-coverage}(e, d) \cdot \textit{e-tf-idf}(e, d)
\end{equation}

\textbf{Graph Cores.}
With our previous scoring functions, we can now determine the relevance of different components of each document graph.
For instance, the statement extraction step might yield multiple edges (with different predicates) between two nodes, e.g., a general \textit{association} and a specific \textit{treats} predicate. 
For our recommendation step, we filter the graphs by only keeping the most relevant (best-scored) edge between two concepts and by only keeping edges between different concept types, e.g., drug-disease treatments. 
A \textbf{graph core} is a scored document graph (i.e., the scoring functions have been applied to each node and edge) that only keeps the best-scored edge between two nodes. 
The function $\textbf{graph-core}(d) = d_{\textit{core}}$ takes a document and returns its core.

\subsection{Candidate Retrieval (First Stage)}
The discovery system contains about 37 million documents (as of 04/2024). 
Given some input document $d$, comparing its core to that of every other document is obviously too expensive.
That is why we headed for a two-step approach: A cheap first stage for initial candidate retrieval and a more expensive second stage that utilizes our graph cores and re-scores the candidate documents. 
Given the document collection $\mathcal{D}$ and some input document $d$, its graph $d_{\textit{graph}}$ and its graph core $d_{core}$, we propose the following first stages:

\textbf{Edge-driven (FSCore).}
The idea for this first stage is that the more edges some document $d_{\textit{candidate}}$ contains with the same concepts in relation as in $d_{\textit{core}}$, the better it is. 
We iterate over each edge $e = (s, p, o)$ of $d_{\textit{core}}$.
We then search for documents containing an edge between $s$ and $o$ (we do not force the edge's predicate to be $p$ to be more flexible). 
We add those documents to a set of candidate documents $D_{\textit{candidate}} \subseteq \mathcal{D}$.
We add the score $\textit{e-score}(e, d)$ to each candidate document that contains the searched edge. 
We continue with our iteration. 
Finally, we retrieve a scored list of documents: The higher the score is, the more overlap it has with edges of $d_{\textit{core}}$.

\textbf{Node-driven (FSNode).}
FSCore requires that the document $d$ has a core (more than zero edges must have been extracted). 
However, due to error-prone NLP methods and extractions on sentence levels (but not across sentences), we might lack recall from the underlying discovery system.
That is why we also introduce a concept-driven first stage.
We apply a similar idea to that in FSCore, but this time, we do not search for edges; instead, we search for nodes.
That means we iterate over the annotated concepts $d_{\textit{concepts}}$, retrieve documents that contain the concept as one of its nodes, and add the concept's score to them.

\textbf{Concept-driven (FSConcept).}
FSNode does not force that the input document $d$ has a graph but searches for candidate documents with $d$'s concepts on their graphs.
As an alternative, we propose FSConcept, which just requires that the searched concepts be detected in the candidate documents but do not necessarily need to be contained on their graphs.
FSConcept allows a concept-driven retrieval without relying on graphs in the input or candidate documents.

\textbf{Cutoff.}
We restrict our first stages by a fixed cutoff value, $k$, so we return only the best-scored $k$ documents.
If the score is equal, we sort documents by their IDs in descending order as our system maintains PubMed IDs and higher IDs usually mean newer publications. 
While BM25 computes nearly continuously distributed scores because it also considers the tf-idf scores of terms within candidate documents, our first stages come with a step function (either a component of the input is contained or not). 
For instance, the documents between rank 900 and 1200 could have the same score, as they contain the same overlap to the input. 
For that, we propose a flexible cutoff which considers the score at position $k$ and then cuts the list at the next position the score drops again. 
However, a flexible cutoff may result in very large lists, that is why, we use a hard cutoff at $2 \cdot k$ in any case, to have a maximum boundary.

\begin{algorithm}[t]
\caption{Recommendation Strategy based on Document Cores}
\label{alg:corerecommendation}
\begin{algorithmic}[1]
\State \textbf{Input}: $d_{input}, D_{candidates}$   \textbf{Output}: a ranked list of documents 
\State  $core_{\textit{input}} = \textit{graph-core}(d_{input})$
\If{$core_{\textit{input}} = \emptyset$}
\State \textbf{return} candidate documents without sorting them
\EndIf
\For{$d_{\textit{candidate}} \in D_{\textit{candidates}}$}
\State score[$d_{\textit{candidate}}$] = 0
\State $core_{\textit{candidate}} = \textit{graph-core}(d_{\textit{candidate}})$
\For{$e \in (core_{\textit{input}} \cap core_{\textit{candidate}})$}
    \State $\textit{score}[d_{\textit{candidate}}] = [d_{\textit{candidate}}] + \textit{e-score}(e, d_{\textit{input}})$
\EndFor
\EndFor
\State \textbf{return} \textit{sorted candidate documents by their scores}

\end{algorithmic}
\end{algorithm}

\subsection{Recommendation (Second Stage)}
The first stage returns a list of candidate documents $D_{\textit{candidates}}$. 
For our second stage, we compute each candidate document's score by comparing their cores to our input document.
Our proposed method is shown in Algorithm~\ref{alg:corerecommendation}.
In brief, if the input document does not have a core, we cannot compute scores and consider all input documents as equally relevant.
If the input document has a core, we compare it to every core of the candidate documents. 
The score for the candidate document is defined as the sum of all edges shared between the input document core and the candidate document's core. 
An edge is considered as \textit{shared} if it connects the same concepts. 
This strategy, however, relies on the existence of cores and the expression of relevant information in these cores. 
This might not always be the case:
First, extraction methods are error-prone, i.e., relevant information might be lost.
Second, our concept vocabulary might not contain \textit{all} relevant concepts of that domain~\cite{DBLP:journals/jodl/KrollPKKRB24}.
That is why we also integrate text-based scoring to consider not-as-graph-expressed information. 
Here, we use BM25 scoring by considering titles and abstracts.
Let $d_i$ be the input document and $d_c$ be some candidate document.
We compute the final score, as a weighted sum of the graph overlap and BM25 score:

\begin{equation}
    \textbf{XGPRec}(d_i, d_c) = w_{\textit{graph}} \cdot \textit{core-overlap}(d_i, d_c) + w_{\textit{text}} \cdot BM25(d_i, d_c)
\end{equation}

The $\textit{core-overlap}$ returns the score of the core overlap based on our previous algorithm, and BM25 returns the BM25 score when comparing the text (title plus abstract) of $d_i$ and $d_c$.
Note that we normalize \textit{core-overlap} and BM25 scores with regard to a candidate document list $D_{\textit{candidate}}$ so documents receive scores between [0, 1] which makes both scores comparable and combineable.

\begin{figure*}[t]
    \centering
    \includegraphics[width=0.7\textwidth]{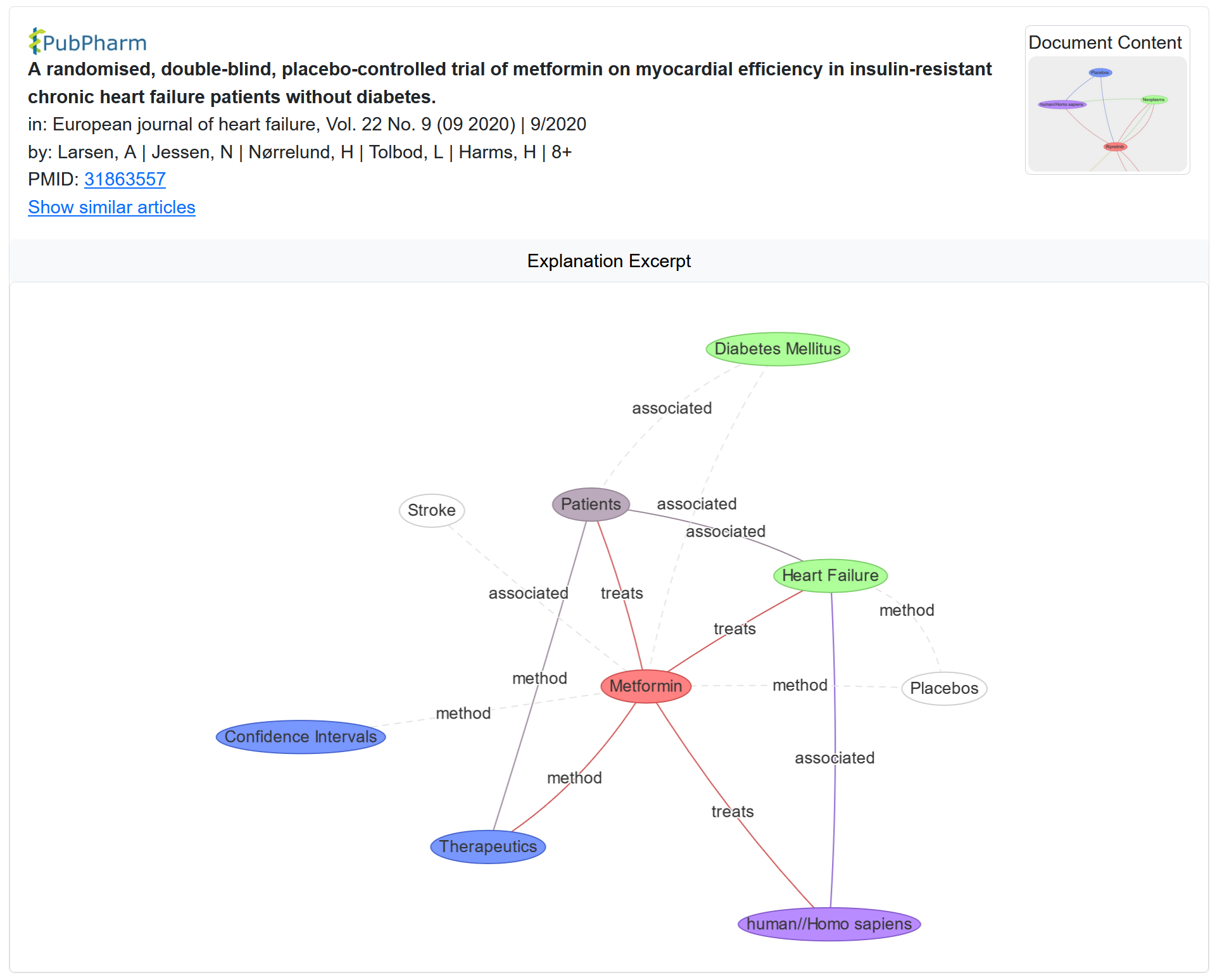}
    \caption{Screenshot of our prototypical system: The generated explanation why the candidate document should be relevant to the input document is shown. Shared information (nodes and edges) are visualized with colors whereas information that is added by the candidate document is visualized as dashed lines and not colored nodes. }
    \Description{Screenshot of our prototypical system: The generated explanation why the candidate document should be relevant to the input document is shown. Shared information (nodes and edges) are visualized with colors whereas information that is added by the candidate document is visualized as dashed lines and not colored nodes. }
    \label{fig:prototype}
\end{figure*}

\begin{figure*}[t]
    \centering
    \includegraphics[width=1.0\textwidth, trim={0cm 2cm 0cm 0cm}]{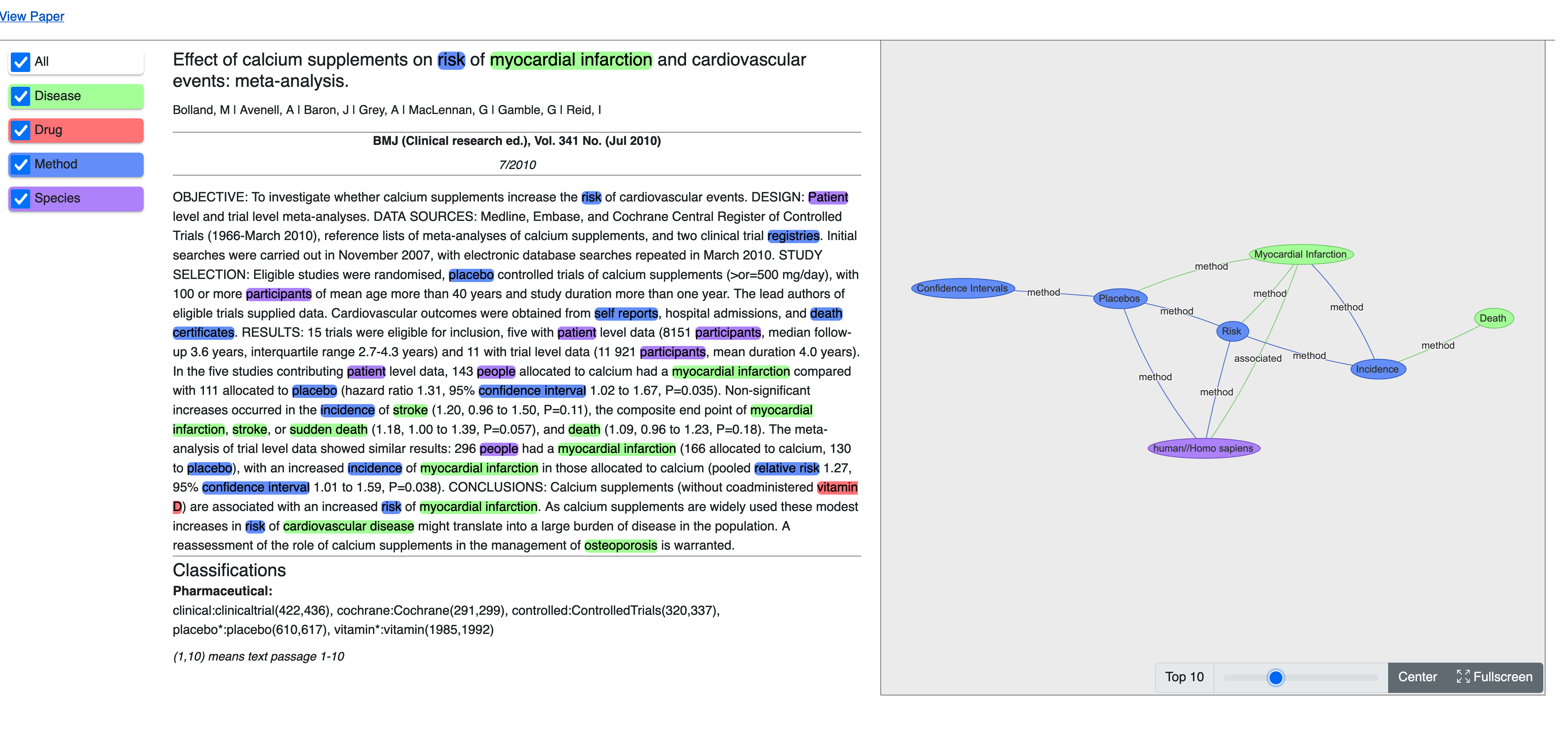}
    \caption{Screenshot of our improved document visualization: On the left side, detected concepts are highlighted in the text via a color encoding. On the right side, the essential document graph is shown, i.e., the most relevant extracted statements are shown to the users. Users can on the left side filter for certain concept types (diseases, drugs, etc.) and on the right side select the number of statements that should be shown. }
    \Description{Screenshot of our improved document visualization: On the left side, detected concepts are highlighted in the text via a color encoding. On the right side, the essential document graph is shown, i.e., the most relevant extracted statements are shown to the users. Users can on the left side filter for certain concept types (diseases, drugs, etc.) and on the right side select the number of statements that should be shown. }
    \label{fig:document_graph}
\end{figure*}

\subsection{Explanation Generation}
Algorithm~\ref{alg:explanation} shows our explanation approach.
The algorithm takes an input document, some candidate document, and a parameter $l$ as its input.
The parameter $l$ determines the length of the explanation to generate. 
Our idea is to take $l$ edges shared between the input documents and the candidate and mark them as shared (later visualized in colors).
In addition, we take up to $l \cdot 2$ edges of the candidate document, mark them as not shared (later visualized as dashed lines and not-colored nodes), and add them to our explanation.
More precisely, we only consider the edges of the candidate core connected to one node shared between the input and candidate core. 
These additional edges should help the user to understand what the candidate documents add as new information to the shared pattern. 
This way users simultaneously see what is shared and what can be expected as new information in the candidate document.

\begin{algorithm}[t]
\caption{Explanation Generation Algorithm}
\label{alg:explanation}
\begin{algorithmic}[1]
\State \textbf{Input}: $d_{\textit{input}}, d_{\textit{candidate}}, l$  \textbf{Output}: A list of shared/not shared edges
\State $core_{\textit{input}} = \textit{graph-core}(d_{input})$
\State $core_{\textit{candidate}} = \textit{graph-core}(d_{candidate})$
\For{$e \in \textit{sort by score}(core_{\textit{input}} \cap core_{\textit{candidate}})$}
\State $\textit{explanation} = \textit{explanation} \cup (e, \textit{shared})$
\If{$|\textit{explanation}| \geq l$}
\State \textbf{break}
\EndIf
\EndFor

\For{$e = (s, p, o) \in \textit{sort by score}(core_{\textit{candidate}} \setminus core_{\textit{input}})$}
\If{$|\textit{explanation}| \geq 2 \cdot l$}
\State \textbf{break}
\EndIf
\If{$s \in \textit{nodes}(core_{\textit{input}})$}
\State $\textit{explanation} = \textit{explanation} \cup (e, \textit{o not shared})$
\EndIf
\If{$o \in \textit{nodes}(core_{\textit{input}})$}
\State $\textit{explanation} = \textit{explanation} \cup (e, \textit{s not shared})$
\EndIf
\EndFor

\State \textbf{return} explanation

\end{algorithmic}
\end{algorithm}

\section{Implementation and Prototype}

We implement our recommendation algorithm by building upon our~\cite{DBLP:journals/jodl/KrollPKKRB24} discovery system's code base.
The discovery system already maintains indexes that can be used to estimate the idf scores for nodes and edges.
More precisely, an index maps a concept to a list of document IDs in which the concept has been detected.
For the idf computation, it is enough to store a dictionary that maps concepts to the number of documents it appears in.
This index contains 1M concepts and takes 40MB of space.
We keep it in memory.
Suppose a user enters a document ID through manual input or via a link to our system. 
In that case, we retrieve the document's data from the database, perform first-stage retrieval, select the $k$ best-scored documents, and then retrieve the actual candidate document data for the recommendation. 
Thus, we load complete document data only if required.

The graph-based discovery system~\cite{DBLP:journals/jodl/KrollPKKRB24}, which we used to implement XGPRec has only utilized concept and graph information of documents.
The system maintains two indexes:
A reverse index for concepts and a reverse index for edges, allowing it to retrieve documents that support a certain concept/edge (retrieve a set of document IDs).
Our first stage can thus reuse both indexes.
Both indexes are stored as database tables and details are available in the service's code repository. 
We ignored 20 very generic concepts like \textit{therapy} or \textit{humans} as they were detected in more than 1M documents and thus carry few information. 
For the BM25 computation, we created a new BM25 index by utilizing PyTerrier~\cite{DBLP:conf/ictir/MacdonaldT20} (a Python Wrapper around the well-known Terrier toolkit).
This index requires about 9.2GB of disk space and can optionally be retrieved from disk or directly from memory.

\textbf{Parameters.} 
For the final XGPRec score, we slightly prefer graph scores over text scores, i.e., we set $w_{\textit{graph}} = 0.6$ and $w_{text} = 0.4$. 
We set predicate specificity scores (see tf-idf score for edges) based on each predicate's hierarchical level in the three-level predicate taxonomy (most-specific predicates received a score of 1.0, one level higher 0.5, and the highest level (only associated) 0.25) defined by the discovery system\footnote{See the taxonomy at \url{https://narrative.pubpharm.de/help/}.}.
We set $k = 1000$.

\textbf{User Interface.}
Figure~\ref{fig:prototype} shows a screenshot of our prototype.
In our user interface the user can enter a document ID as an input.
Then, a list of candidate documents is retrieved and ranked via our recommendation strategy. 
For each entry of that list, we generate an explanation and visualize it as shown in Figure~\ref{fig:prototype}.
We used the same color encoding used in the Narrative Service to visualize document graphs (e.g., red for drugs, green for diseases~\cite{DBLP:journals/jodl/KrollPKKRB24}).
Our graph patterns should help users quickly determine the information scent of the recommendation list -- a feature not available in other systems. 
For the explanation visualization, we tested different $l$ values (no. of shown edges when generating an explanation). 
We found that six (six edges of the core overlap and six additional edges from the candidate document) generates a graph pattern that fits into the user interface concerning the available space. 
We also improved our old document graph visualization~\cite{DBLP:journals/jodl/KrollPKKRB24}.
The old system basically showed all extracted statements as one graph.
Our improved visualization now shows document graphs that are reduced to the most essential parts by edge scoring, which should decrease the complexity for users.
However, users can still control which concept types and thus, which interactions between concept types, are shown as well as the number of statements to show (default are ten statements). 
Our visualization is shown in Figure~\ref{fig:document_graph}.

\section{Evaluation}
We (1) describe the used test collections, (2) evaluate the recall and runtime of our first stage, (3) report the precision and recall of our recommendation strategy, and (4) close with a human perspective of our prototypical system.

\begin{table*}[t]
\caption{Evaluation of our first stages: The set recall and average retrieval time per document is shown. We used a cutoff $k = 1000$. The flexible cutoff is reported as $^F$.}
\centering
\begin{tabular}{l|@{\hskip 0.15cm}c@{\hskip 0.15cm}c|@{\hskip 0.15cm}c@{\hskip 0.15cm}c|@{\hskip 0.15cm}c@{\hskip 0.15cm}c}
\toprule
\textbf{Ranking} & \multicolumn{2}{c}{\textbf{PM2020}} & \multicolumn{2}{c}{\textbf{Genomics}} & \multicolumn{2}{c}{\textbf{RELISH}} \\

& Recall& $T_{doc}$ & Recall& $T_{doc}$ & Recall& $T_{doc}$ \\
\midrule
FSCore & 0.52 $\mid 0.53^F$ & 1.3s & 0.22  $\mid0.22^F$  & 0.1s & 0.16 $\mid 0.17^F$ & 0.3s \\
FSNode& 0.60 $\mid \textbf{0.61}^F$ &  0.8s & 0.27 $\mid 0.29^F$ &  0.1s & 0.22 $\mid 0.24^F$ & 0.4s\\
FSConcept & 0.60 $\mid \textbf{0.61}^F$ & 0.9s & 0.27 $\mid 0.29^F$ & 0.1s & 0.23 $\mid 0.25^F$& 0.5s\\

\midrule
BM25 Title & 0.50 & 1.1s & 0.33 &  0.2s & 0.61 & 0.6s \\
BM25 Title + Abstract & 0.57  &  9.3s & \textbf{0.41} & 1.2s & \textbf{0.80} & 8.3s \\

\bottomrule
\end{tabular}

\label{tab:first_stage}
\end{table*}

\subsection{Test Collections and Baselines}
We use the following test collections:

\textbf{TREC Genomics 2005} (Genomics)~\cite{DBLP:conf/trec/HershCYBRH05}: (50 topics/2525 input documents) contains natural language questions about genes, interactions, processes and methods for ad-hoc biomedical document retrieval. The information retrieval test collection was built with the MEDLINE 2005 (3.7M documents). 
It has already been used in previous paper recommendation evaluations~\cite{DBLP:conf/medinfo/GuoSZH21,DBLP:journals/bmcbi/LinW07,DBLP:journals/jbi/ZhangLCHC22}.
We follow Zhang et al.~\cite{DBLP:journals/jbi/ZhangLCHC22} in selecting relevant articles (judged as 2 - relevant) per topic as input documents while considering all other articles belonging to the same topic as potential candidate documents with their corresponding judgments (2 - relevant/1 - partially relevant/0 - not relevant).
We used the default configuration of trec\_eval to compute the scores. 
Then, we iterate over all relevant (2) articles of a topic, perform the recommendation, compute scores, and then average the scores for all input documents per topic. 
In this way, we derive an average score per topic by iterating over all combinations (input vs. documents to recommend). 

\textbf{TREC Precision Medicine 2020} (PM2020)~\cite{TREC_PM20}: (31 topics/1192 input documents) is a biomedical document retrieval test collection that asks for treatment options (drug), cancer forms (disease), and a gene variant (gene/target). 
The collection was built with the MEDLINE 2019 (29.1M documents).
We performed our evaluation analogous to Genomics. 

\textbf{RELISH}~\cite{DBLP:journals/biodb/BrownCZ19}: (3278 input document queries) is a biomedical paper recommendation test collection with 161184 distinct documents which have been rated by experts. It was built with the MEDLINE 2018 (26.7M documents).

\textbf{Data Processing.}
Our discovery system~\cite{DBLP:journals/jodl/KrollPKKRB24} already included the latest version of MEDLINE, so the benchmark documents have already been processed and transformed into document graph representations.
For our evaluation, we computed a list of document IDs belonging to the different MEDLINE versions (2005, 2018, and 2019). 
This list was used as a filter to retrieve only documents in our first stages that were initially considered in the benchmarks.

\textbf{Baselines.}
We compare our first stages to BM25. 
Related work rarely publishes their code or describes their methodology in enough detail to enable reimplementation~\cite{DBLP:journals/jodl/KreutzS22}.
Therefore, we decided to compare against a method applicable to real-world data, that is a pre-existing system, namely PubMed~\cite{DBLP:journals/bmcbi/LinW07} (see Sec.~\ref{sec:related_work}).

As another configuration for the second stage we compared XGPRec to the documents retrieved by BM25 on titles and abstracts (the best first stage) reranked with SPLADE~\cite{DBLP:conf/sigir/FormalPC21}. SPLADE uses lexical matching and term expansion for the embedding of query and documents and achieved results on par with state-of-the-art dense retrieval models. 
Similarity between SPLADE-embedded\footnote{By the \href{https://huggingface.co/naver/splade-cocondenser-ensembledistil}{naver/splade-cocondenser-ensembledistil} model and max pooling.} candidate document vectors and query documents is computed with the Faiss~\cite{douze2024faiss,johnson2019billion} L2 norm. We then identify the $k$ documents to recommend by applying knn.

\begin{table*}[t]
\caption{Detailed evaluation of our recommendation approach XGPRec.}
\centering
\begin{tabular}{l@{\hskip 0.2cm}l@{\hskip 0.2cm}c@{\hskip 0.2cm}cccccc}
\toprule

Dataset & Strategy & $T_{doc}$ &Recall & nDCG@10 & nDCG@20 & P@10 & P@20 & bpref \\
\midrule
PM2020 & \textbf{XGPRec} & $0.45 \pm 0.38s$ & \textbf{0.61} & 0.30 & 0.30 & 0.33 & 0.30 & 0.30 \\
& \quad - BM25 & $0.44 \pm 0.38s$ &\textbf{0.61} & 0.25 & 0.25 & 0.28 & 0.26 & 0.29 \\
& \quad - CoreOverlap  & $\le 0.1 \pm 0.0s$ & \textbf{0.61} & \textbf{0.32} & \textbf{0.31} & \textbf{0.34} & \textbf{0.30} & \textbf{0.31} \\
\midrule
& BM25 Title & 1.1s & 0.50 & 0.23 & 0.23 & 0.25 & 0.23 & 0.25 \\
& BM25 Title + Abstract & 9.3s & 0.57 & 0.29 & 0.28 & 0.31 & 0.28 & 0.28 \\
& BM25 T+A + SPLADE   & - & {0.57} & 0.22 & 0.2 & 0.23 & 0.2 & 0.25 \\
\midrule
& PubMed Rec. & - & 0.29 & 0.30 & 0.30 & 0.33 & 0.29 & 0.17 \\

\midrule
\midrule
Genomics & \textbf{XGPRec} & $0.13 \pm 0.18s$  & {0.29} & 0.20 & 0.19 & 0.21 & 0.18 & 0.15 \\
& \quad - BM25 & $0.13 \pm 0.18s$ & {0.29} & 0.14 & 0.13 & 0.16 & 0.13 & 0.14 \\
& \quad - CoreOverlap & $\le 0.1 \pm 0.0s$ & {0.29} & {0.24} & {0.22} & {0.25} & {0.21} & {0.16}\\

\midrule
& BM25 Title & 0.2s & 0.33 & 0.19 & 0.18 & 0.21 & 0.18 & 0.17 \\
& BM25 Title + Abstract & 1.2s &  \textbf{0.41} & \textbf{0.26} & \textbf{0.25} & \textbf{0.28} & \textbf{0.24} & \textbf{0.21} \\
& BM25 T+A + SPLADE  & - & \textbf{0.41} & 0.19 & 0.18 & 0.20 & 0.17 & 0.17 \\
\midrule
& PubMed Rec. & - & 0.13 & 0.23 & {0.22} & 0.24 & 0.20 & 0.09 \\

\midrule
\midrule
RELISH &\textbf{XGPRec} &$0.10 \pm 0.33s$ &0.25 & 0.22 & 0.19 & 0.22 & 0.17 & 0.22 \\
&\quad - BM25 & $0.10 \pm 0.33s$& 0.25 & 0.09 & 0.08 & 0.09 & 0.07 & 0.21 \\
&\quad - CoreOverlap & $\le 0.1 \pm 0.0s$ &  0.25 & 0.35 & 0.29 & 0.35 & 0.26 & 0.22 \\

\midrule
& BM25 Title & 0.6s & 0.61 & 0.33 & 0.29 & 0.33 & 0.27 & 0.47 \\
& BM25 Title + Abstract & 8.3s & \textbf{0.80} & 0.54 & 0.49 & 0.57 & 0.47 & 0.59 \\
& BM25 T+A + SPLADE  & - & \textbf{0.80} & 0.29 & 0.26 & 0.30 & 0.23 & 0.57 \\
\midrule
& PubMed Rec. & - & {0.47} & \textbf{0.55} & \textbf{0.51} & \textbf{0.60} & \textbf{0.52}  & \textbf{0.41} \\

\bottomrule
\end{tabular}
\label{tab:recommendation}
\end{table*}

\subsection{First Stage Evaluation}
Table~\ref{tab:first_stage} shows the recall and average retrieval time per document for different first stages, i.e., FSCore, FSNode, FSConcept, BM25 with titles (BM25 T) and BM25 with titles and abstracts (BM25 T+A). 
We measured the runtime performance on our server, which has two Intel(R) Xeon(R) Gold 6336Y CPUs @ 2.40GHz (24 cores and 48 threads each), 2TB DDR4 main memory, and nine Nvidia A40 GPUs with 48GB memory. 
All time measurements were executed four times. 
The first run was a cold start to load all required data into main memory, and the reported results were averaged over the last three runs.
All data for BM25 was also loaded into main memory, and no multi-threading/processing was used. 

In concept-centric cases like PM2020, FSConcept, and FSNode achieved the highest recall of 0.6-0.61 and the lowest retrieval time of 0.8-0.9s per input document.
In comparison, BM25 T+A achieved a recall of 0.57 by requiring about 9.3s per input document.
On Genomics, our methods achieved a recall comparable to BM25 T (-0.04) with a retrieval time comparable to BM25 T (0.1s vs. 0.2s). 
BM25 T+A achieved a recall of 0.41 but required about 1.2s per document.
Please note that these times are measured by using a BM25 index of only 3.7M documents, compared to 9.3s (BM25 T+A) on 29.1M documents in PM2020.  
On RELISH, BM25 T and BM25 T+A outperformed our first stages' recall.
In brief, the runtime for our first stages are always below BM25 retrieval and in concept-centric scenarios, our first stages are effective (PM2020) or comparable (Genomics).  
We discuss shortcomings on RELISH in Section~\ref{sec:benchmark_discussion}.

\subsection{Recommender Evaluation}
Our first stages FSConcept and FSNode showed a comparable performance. 
We decided to use FSConcept with a flexible cutoff as the first stage for our recommendation approach because (1) we do not need an additional index (as the node graph index for FSNode) and (2) FSConcept is less restrictive than FSNode (it just requires concepts to be annotated in documents and not that these concepts need to appear on the document graph).
Table~\ref{tab:recommendation} shows the results of our recommendation approach (XGPRec) compared to the PubMed Recommender. 
We also filtered the PubMed Recommender results so that only documents are used for the evaluation considered in the test collections. 
In addition, we report the results of using XGPRec without the core overlap score (-CoreOverlap) and without the BM25 score (-BM25).
In general, our recommendation strategies took less than 1s per document for the computation (see $T_{doc}$ in Table~\ref{tab:recommendation}).
On PM2020 and Genomics, XGPRec achieved comparable nDCG and precision to the PubMed Recommender but nearly doubled recall (0.61 vs. 0.29 and 0.29 vs. 0.13). 
However, PubMed's lower recall could be explained by smaller result sets (maybe because of some internal precision-oriented cutting strategy). 
On RELISH, however, our method was outperformed by the PubMed Recommender.
Another observation was that XGPRec without the CoreOverlap component achieved the highest scores, i.e., by just using the BM25 scores. 

In comparison, using only BM25 on titles or titles and abstracts of input documents instead of employing a recommender system achieves good results on the three benchmarks\footnote{Note that both PM2020 and Genomics are retrieval benchmarks and that BM25 is a retrieval method. These benchmarks were built by annotation of documents which have been highly ranked by retrieval systems, see Sec.~\ref{sec:benchmark_discussion}.}. The variant using abstracts unsurprisingly produces higher recall, nDCG, precision and bpref than the one using titles only. The comparably high execution time makes this strategy uneligible in a real system. Results produced by BM25 title are comparable to XGPRec except for the RELISH case.

For the combination of BM25 on titles and abstracts (BM25 T+A) as the first stage and using SPLADE and Faiss for reranking in the second stage we do not report recommendation times. SPLADE embeddings can theoretically be pre-computed for the whole document corpus\footnote{Which we refrained from doing as we would have been faced with embedding 38M documents which was not feasible for our digital library use case.} thus our measured times would not reflect the actual optimized usage options of Faiss (see ~\cite{johnson2019billion,douze2024faiss}).
The BM25 T+A + SPLADE run produces results for nDCG, precision and bpref that are considerably worse than XGPRec on PM2020, similar on Genomics and considerably better on RELISH. As this variant is a reranking of the results of BM25 T+A, their recall does not differ.

\subsection{Differences between XGPRec and PubMed}

\begin{table}[t]
\caption{Differences between XGPRec, BM25 and the PubMed recommender. J@$k$ gives the average Jaccard coefficient over the sets of documents at top $k$ recommendations for the three benchmarks PM2020, Genomics, and RELISH.}
\centering
\begin{tabular}{l@{\hskip 0.2cm}l@{\hskip 0.2cm}l@{\hskip 0.2cm}cccc}
\toprule
XGPRec vs. & Dataset  & J@5 & J@10 & J@20 & J@100 \\
\midrule
BM25 Title+Abstract & PM2020 & 0.24 & 0.25 & 0.26 & 0.30 \\
& Genomics & 0.23 & 0.23 & 0.23 & 0.26 \\
& RELISH & 0.10 & 0.10 & 0.10 & 0.09 \\
\midrule
PubMed & PM2020 & 0.13 & 0.14 & 0.16 & 0.18  \\
& Genomics &  0.13 & 0.14 & 0.15 & 0.13 \\
& RELISH & 0.08 & 0.07 & 0.07 & 0.07 \\
\bottomrule
\end{tabular}
\label{tab:differences}
\end{table}

Table~\ref{tab:differences} contains average Jaccard coefficients of sets of documents from top $k$ recommendation comparing XGPRec with BM25 using titles and abstracts as well as XGPRec with the PubMed recommender on our benchmarks. The overlap in documents is similar between different $k$ per row, that is between benchmark and comparison. The overlaps between PM2020 and Genomics are comparable as well. The overlaps between recommended documents using XGPRec and the BM25 strategy as well as PubMed is considerably lower in the RELISH dataset. In general, XGPRec recommends documents more similar to the BM25 baseline than PubMed.
Recommendations made by XGPRec are dissimilar from those made by the PubMed recommender.
This dissimilarity can be regarded with the comparable performance of XGPRec and PubMed on the PM2020 and Genomics benchmarks to conclude that XGPRec presents documents to its users which they would not be recommended using PubMed but which are of similar relevance, as estimated by XGPRec. We therefore argue that XGPRec holds value in helping users to uncover relevant documents which the state-of-the-art in-use recommendation system does not suggest.

\subsection{Result Discussion}
\label{sec:benchmark_discussion}
Usually, collections are crafted by using some initial pooling strategy of candidate documents for each topic and then showing them to judges who rate whether the document is relevant.
RELISH was constructed by using BM25, tf-idf and the PubMed recommender for initial pooling~\cite{DBLP:journals/biodb/BrownCZ19}. 
That explains why the PubMed recommender shows good performance on RELISH.
In general, when term-based retrieval is used for an initial pooling, term-based methods like BM25 are automatically favored. 
Advantages of graph-based retrieval, e.g., synonymous terms that refer to the same concept, are then rather useless than helpful. 
Especially, when documents that are not judged in the benchmarks are counted as wrong hits. 

While some paper recommendation methods \textit{just} divide relevant and irrelevant documents given by some test collection, as done in \cite{DBLP:journals/biodb/BrownCZ19,10.1093/bioinformatics/btad651,RavinderFellerhofDadietal.2023,DBLP:journals/jbi/ZhangLCHC22}, our system works on a comprehensive document collection. 
When evaluating the test collections, we observed two central issues:
First, our graph-based approach retrieved many documents that have not been judged in the collection data.
Those documents are then considered as incorrect results which hampers our nDCG and precision@k scores.
However, the metric bpref~\cite{Craswell2009}, which ignores unjudged documents, shows that the difference between core overlap and BM25 is small (1-2\% points).
In fact, under the top-20-retrieved documents with XGPRec, we observed $6.2 \pm 5.8$ unjudged documents on average per topic for PM2020, $9.8 \pm 6.3$ for Genomics, and $15.2 \pm 5.4$ for RELISH.
The number of unjudged documents decreased for XGPRec without CoreOverlap and only BM25 scores: $5.9 \pm 5.6$ for PM2020, $9.0 \pm 6.2$ for Genomics, and $12.9 \pm 5.9$ for RELISH. 
Second, not all test collection documents can be represented by a document graph and salient concepts; hence, our method cannot retrieve documents in those instances.
We counted the ratio of input documents that have a core with five edges. 
This is the case for 96.0\% documents for PM2020, 83.0\% for Genomics, and 55.0\% for RELISH. 
Thus, we clearly lacked suitable document graph representations on RELISH.
The discovery system has been built for the pharmaceutical domain, i.e., mainly for gene, drug, and disease interactions (reflected in PM2020 and Genomics) and not for psychological or orthopedic topics as present in RELISH.  

While both issues explain that BM25 achieved higher scores than our graph-based overlap method, our graph method has the advantage of preferring overlapping graphs. 
Moreover, these overlapping graphs can explain recommendations for users. 
We already know domain experts like using graphs in the context of literature search to quickly determine if results fit their information needs~\cite{DBLP:conf/jcdl/KrollKSB23}.

While PubMed and XGPRec produce comparable results on PM2020 and Genomics, the actually recommended documents are vastly different from each other. Therefore, we argue XGPRec being a suitable tool to uncover potentially relevant documents which PubMed does not suggest to users. We consider XGPRec not as a \textbf{better} recommendation system compared to PubMed for producing relevant recommendations but instead want to emphasize on XGPRec's value to help discover \textbf{different} documents in an explainable manner.

In brief, XGPRec offers explanations by design, can handle a real-world digital library collection with about 37M documents, and offers a comparable performance to the PubMed recommendation (which is likely the most used recommender in the biomedical domain). Beyond that, XGPRec retrieves result lists that differ from existing benchmarks (results are not judged there) and recommendation approaches. These results are worth of further investigation. 

\subsection{Human Perspective}
To get initial user feedback on our recommendation system we separately interviewed two experts from the pharmaceutical domain for 30 minutes. The experts were also knowledgeable users of the underlying discovery system. An interviewer first briefly described the interface's functionality, before the domain experts used the system for 20 minutes to satisfy information needs while thinking aloud. Afterwards the interviewer asked some clarifying questions regarding liked or disliked components, missed functionality and general feedback.

\textbf{Results.}
Both users looked at five (individually chosen) initial papers and commented on their recommendations' relevance. They both searched for 1) interactions or 2) keywords contained in an initial paper. They both observed broader and more focused information needs.
Our users enjoyed using the system, were very positive about the quality of recommended papers and especially praised the explanation graphs. They were immediately able to use the visualization of the overlap between their initial paper and the candidate suggestions. They were able to interpret the graphs at a glance and distinguish relevant from less relevant suggestions instantly without having to read the title of the papers.
One participant only observed the graphs of recommendations and did not click a single detail view while the other preferred to first read titles and if titles were hinting towards relevance, they checked out the graphs. When first screening titles, it would be desirable to have the graphs less prominent.
It was mentioned that the graph was considered a better option for screening papers than a keyword list.
One person used the option of observing similar articles of recommended articles, thus chaining recommendations to navigate the content of the underlying digital library.

Features which participants missed were a structuring of the result list, e.g., by types of overlapping edges/patterns and the possibility to emphasize/weight edges from input papers which a user considers highly relevant.
Purely technical wished for functionality were the option to keep the initial paper displayed on the same page while checking the recommendations, a bookmarking or export option for relevant articles, filtering options on the result list, and a description how the recommendation system determines the suggestions, i.e., a brief description of our used method.

\textbf{Discussion.}
Our interview partners used the graph overlap between the initial and a candidate paper intuitively to understand, why a paper was recommended. The explanation therefore satisfies the aims of justification~\cite{DBLP:conf/sigir/BalogRA19}.
One study participant mentioned the current interface lacking information how our systems generates the list of recommendations~\cite{DBLP:conf/sigir/BalogRA19} which should be tackled by some info box. 
Chaining recommendations seems to be a good way of exploring a topic. 
Even though our evaluation with test collections did not produce the best results numerically, our preliminary user study showed participants' immediate acceptance of the system and their satisfaction with the recommendations. 
This could hint at a mismatch between test collections and the human perspective in paper recommendation - collections are absolutely needed to construct systems but they must be complemented with a user study to further shed light on advantages and limitations of systems.

As a next step in development and after confirming the user interface's preliminary suitability, we intend to bring the system into a beta phase and run a large scale study in order to explore explanation options for recommendations.

\section{Conclusion}
This work extended our graph-based discovery system by an explainable paper recommendation component for a real-world digital library document collection. 
In contrast to many other works, our method (1) is unsupervised, i.e., we do not require training data and a library must thus not collect training data to implement a similar algorithm, and (2) it works on a real, large-scale collection with 37M documents.
While our evaluation shows benefits and limitations, we demonstrated an overall comparable performance to the PubMed recommender while suggesting a vastly different set of documents and therefore potentially broadening a user's exploration space.
We argue that precise, concept-centric information needs are common in the biomedical domain, as seen in a PubMed query log analysis~\cite{herskovic2007pubmedqueryanalysis}, PM2020~\cite{TREC_PM20}, our previous user study~\cite{DBLP:journals/jodl/KrollPKKRB24}, or our discovery system's query log analysis~\cite{DBLP:conf/jcdl/KrollKSB23}.
For these concept-centric use cases like in PM2020 or Genomics, our system shows its advantages: 
The first stage is fast and effective and the core overlap recommendation allows to derive graph pattern explanations which help our users.
In brief, our recommendation strategy XGPRec is fast in handling an extensive collection, offers a comparable performance to PubMed's real digital library recommendation system, and provides users with suitable graph explanations.
This research demonstrates how graph-based document representations allow beneficial exploration in digital libraries.
Beyond that, our code is freely available so that other digital libraries can use or adapt our effective graph-based recommendation implementation.

Explanations must be a focus for future work, i.e., to further improve the visualization strategy by adjusting it to possible user needs. Beyond that, combining graph-based explainable with traditional recommendation methods as well as a user-controlled weighting of graph components are worth of further investigation.

\section*{Acknowledgments}
Supported by the Deutsche Forschungsgemeinschaft (DFG, German Research Foundation): PubPharm – the Specialized Information Service for Pharmacy (Gepris 267140244).

\bibliographystyle{ACM-Reference-Format}
\bibliography{bib}

\end{document}